\documentclass[preprint,floats,aps,superscriptaddress,showpacs]{revtex4}
\usepackage{amsmath}
\usepackage{graphicx}
\newcommand{\be}{\begin{equation}}
\newcommand{\ee}{\end{equation}}
\newcommand{\ba}{\begin{eqnarray}}
\newcommand{\ea}{\end{eqnarray}}

\newcommand{\br}{{\bf r}}

\newcommand{\bv}{{\bf v}}

\begin{document}
\date{\today}

\title{ {\bf Bistable Clustering in Driven Granular Mixtures}}
\author{Giulio Costantini}
\affiliation{Dipartimento di Fisica, Universit\`a di Camerino and
Istituto Nazionale di Fisica della Materia,
Via Madonna delle Carceri, 62032 Camerino, Italy}

\author{Daniela Paolotti}
\affiliation{Dipartimento di Fisica, Universit\`a di Camerino and
Istituto Nazionale di Fisica della Materia,
Via Madonna delle Carceri, 62032 Camerino, Italy}
\author{Ciro Cattuto}
\affiliation{Frontier Research System, The Institute of Physical and Chemical
Research (RIKEN), Wako-shi, Saitama 351-0198, Japan}
\author{Umberto Marini Bettolo Marconi}
\affiliation{Dipartimento di Fisica, Universit\`a di Camerino and
Istituto Nazionale di Fisica della Materia,
Via Madonna delle Carceri, 62032 Camerino, Italy}


\begin{abstract}
The behavior of a bidisperse inelastic gas vertically shaken 
in a compartmentalized container is investigated
using two different approaches: the first is
a mean-field dynamical model, which treats 
the number of particles in the two compartments and the associated 
kinetic temperatures in a self-consistent fashion;
the second is an event-driven numerical simulation. 
Both approaches reveal a non-stationary regime, which has no counterpart in the
case of monodisperse granular gases. Specifically, when the mass difference
between the two species exceeds a certain threshold the populations
display a bistable behavior, with particles of each species switching
back and forth between compartments.
The reason for such an unexpected behavior is attributed
to the interplay of kinetic energy non-equipartition due to inelasticity
with the energy redistribution induced by collisions.
The mean-field model and numerical simulation are found to 
agree qualitatively.
\end{abstract}
\pacs{02.50.Ey, 05.20.Dd, 81.05.Rm}
\maketitle


\section{Introduction}
Granular fluids are currently attracting growing interest in view of their
unusual properties, some of which are not fully understood yet.
Moreover, new experiments continue to reveal
unexpected phenomena, which have no counterpart in molecular fluids.
Clustering, shear instability, non-Maxwellian velocity distributions,
long range velocity correlations, non-equipartition in a binary mixture
are just a few of these peculiarities \cite{gases,general,Kadanoff}. 
Recently, another fascinating phenomenon was reported. It is the so called
``Maxwell sand daemon'' experiment \cite{experiment}, where a system
consisting of inelastic particles enclosed in a two-compartment container
is shaken vertically. 
Particles can flow from one compartment to the other 
through a small orifice located at a certain height from the basal
vibrating plate.
For strong shaking, the right and left populations 
are statistically equal, whereas for weak shaking the system spontaneously
breaks the left-right symmetry.
The mechanism behind such an unusual ordering process is
the clustering induced by inelasticity.

In fact, an imbalance in populations induced by a fluctuation can be
amplified, since it causes a larger energy dissipation on the overpopulated
side, thus suppressing the outflow from that compartment.
At the same time, inflow from the underpopulated compartment is enhanced
because of its lower occupation, which results in higher kinetic energies
per particle \cite{Eggers}.

Quite recently, the Twente collaboration \cite{Mikkelsen} 
investigated the behavior of a bidisperse granular mixture
of small and large particles using a similar experimental setup.
Their experiments demonstrated that a
bidisperse compartmentalized granular mixture has a tendency to
cluster competitively.
Depending on the shaking strength, one can observe different asymptotic
configurations.

Theoretical treatments of 
granular gases in compartmentalized systems range from phenomenological
flux models \cite{Eggers,Mikkelsen,Lohse,Lohse1,Lohse2,Droz,Droz1,Droz2},
to molecular dynamics~\cite{dererumpallettarum,ceccojcp}, 
to more refined kinetic approaches \cite{Brey}.
Whereas the full solution of the inelastic Boltzmann equation
remains a formidable task, a simple set of mean-field dynamical equations
can be derived \cite{Puglio,Conti}.
According to this method, the kinetic temperatures
and the occupation numbers in each compartment are assumed to be
the only relevant dynamical variables and treated  
on equal footing, a technique which naturally lends itself
to capture the more complex phenomenology expected in mixtures.

Here we study a binary mixture of inelastic hard disks in a two-compartment
system. The two species have different masses, are subjected
to gravity and driven by a vibrating base. 
We extend the mean-field treatment of ref.\cite{Puglio} and compare
its predictions with the results of even-driven simulations.

The paper is organized as follows: after introducing our model in section II,
we develop our mean-field treatment in section III.
In section IV we report the predictions of our model and summarize
our findings with a mean-field ``phase diagram'', where the boundaries
between different regimes are studied as function of the control parameters.
In section V we turn to the event-driven simulation and find qualitative
agreement with the previous picture. Finally, in section VI, we present
our conclusions.

\section{The model}
Let us consider a two-dimensional rectangular container with horizontal 
and vertical sides of length $L_x$ and $L_z$, respectively,
divided in two equal compartments by a wall of height $h<L_z$.
The box contains a mixture of $\mathcal{N}_1$ and $\mathcal{N}_2$ inelastic hard disks
of diameter $\sigma$ and masses $m_1$ and $m_2$, respectively. 
The disks are subjected to a gravitational force acting along the negative
$z$ direction and are fluidized by the sawtooth-like movement of the base,
which oscillates with frequency $\nu$ and amplitude $A$. 
The side and top walls are fixed, and particles collide with them
in a perfectly elastic fashion.
The collisions between particles are inelastic, and 
will be described by means of a velocity-dependent coefficient of restitution
\cite{Daniela}:
$\alpha(V_n)$,
\begin{displaymath}
\alpha(V_n) = \left\{
\begin{array}{lll}
1 - (1 - r) \left( \frac{|V_n|}{v_0} \right)^\frac{3}{4} & \mbox{
for } & |V_n| < v_0 \\
      r & \mbox{ for } & |V_n| > v_0 \,\, ,
\end{array}
\right.
\end{displaymath}
where $V_n$ is the pre-collisional relative velocity along the direction
joining the centers of the two particles, $v_0 = \sqrt{g \sigma}$
is a cutoff velocity and $g$ is the gravitational acceleration.
For large values of relative normal velocities, the function $\alpha(V_n)$
assumes a constant value $r$, while for $V_n \to 0$ the elastic behavior
$\alpha(0)=1$ is approached according to a power law.
We checked the robustness of our results with respect to variations in $v_0$.
For the sake of simplicity, the law $\alpha(V_n)$ is the same
for all types of collision.
Binary collision between a particle of species $\mu$ and another particle
of species $\nu$ change 
pre-collisional velocities $\bf{v}_1$ and $\bf{v}_2$
into post-collisional velocities $\bf{v}_1'$ and $\bf{v}_2'$ according to: 
\begin{eqnarray}
{\bf v}_1'= { \bf v}_1 - \mu_{\nu\mu} (1+\alpha)
(\widehat{\bf{n}}\cdot {\bf v}_{12})\widehat{{\bf n}} \\
{ \bf v}_2'= {\bf v}_2 + \mu_{\mu\nu} (1+\alpha)
(\widehat{\bf{n}}\cdot {\bf v}_{12})\widehat{{\bf n}}
\label{trasforma}
\end{eqnarray}  
where $\mu_{\mu\nu}=m_{\mu}/(m_{\mu}+m_{\nu})$,
$\widehat{\bf{n}}$ is a unit vector directed from the center of the particle of type $\mu$ to the
center of particle $\nu$,
and $\bf{v}_{12}= \bf{v}_1 - \bf{v}_2$.
Inter-particle collisions conserve momentum, but result in an energy
loss proportional to $(1-\alpha^2)$.

 The space of the control parameters is large since
the system properties are functions of several dimensionless quantities,
such as the coefficient of restitution $r$, the total number of particles
$\mathcal{N}=\mathcal{N}_1+\mathcal{N}_2$,
the mass ratio, the ratio of lengths $L_x/\sigma$,
the ratio between the typical energy transferred from the piston to the
particles and their gravitational energy, and so on.
We study the system behavior by considering a
combination of the above parameters,
\begin{equation}
R=\frac{ g h}{A^2 \nu^2 }(1-r^2) \frac{\mathcal{N}^2 \sigma^2}{L_x^2} \, ,
\label{R}
\end{equation}
whose relevance has been pointed out in \cite{Lohse1}.
The dimensionless parameter $R$ decreases as the driving intensity
$A\nu$ increases, while increases as the dissipation becomes larger.



\section{Mean-Field Theory}
In this section we shall extend the mean-field treatment
of the compartmentalized inelastic gas, which was introduced 
in refs.~\cite{Puglio,Conti}, to include the case of
different species,
non-equal occupation numbers and different partial temperatures of the system.
In a realistic scenario, the presence of strong inhomogeneity
and of the ensuing gradients make the analysis of the Boltzmann transport
equation far too complex.
The Twente collaboration proposed a simple flux model \cite{Lohse},
while in ref.~\cite{Puglio} a simple coarse-grained version
of the Boltzmann equation was introduced, leading to equations
for the occupation numbers that appear to be very similar to those
of the flux model. Such an approach has also the advantage of treating
occupation numbers and granular temperatures on equal footing.
In the following, we extend this method to the case of a
granular mixture.

In order to obtain our mean-field equations we shall make a series
of simplifying assumptions:
\begin{enumerate}
\item 
the system properties are regarded as homogeneous 
within each compartment, so that only the numbers of particles 
and the kinetic energy of the two species in each compartment
are the only relevant variables;
\item the coefficient of restitution is velocity-independent, i.e.
$\alpha(V_n)=r$;
\item
the driving mechanism is spatially homogeneous and representable
by means of a stochastic thermostat;
\item
the velocity distributions are Gaussian, and their variances
are derived in a self-consistent fashion;
\item
the effect of collisions is adequately described by the inelastic homogeneous
Boltzmann equation; 
\item
particles whose kinetic energy exceeds an assigned threshold
switch to the adjacent compartment at a fixed rate $\tau_s^{-1}$.
\end{enumerate}
The assumption of spatial homogeneity allows us to write the
phase-space distribution function for species $\mu$ (with $\mu=1,2$) as
$$
f_{\mu}({\bf r},{ \bf v},t) = f_{a,\mu}(\bv,t) \,\, .
$$
Here $a = 1$ when point ${\bf r}$ falls within the left side of the container
(relative to the barrier) and $a = 2$ when ${\bf r}$ falls on the right side.
Such a coarse-graining procedure allows us to write 
the following Boltzmann-like equation:
\begin{equation}
\partial_t f_{a,\mu} ( {\bf v}, t) = \sum_{\nu=1}^2 I_{\mu\nu}[ {\bf v} 
| f_{a,\mu}, f_{a,\nu}]+{\cal B}f_{a,\mu}+
{\cal X}[{\bf v}| f_{a,\mu},f_{b,\mu}] \,\, ,
\label{eq:bolt}
\end{equation}
where $I_{\mu\nu}$ is the collision term which describes
the effect of inelastic collisions among particles
belonging to the same compartment, ${\cal B}f_{a,\mu}$
represents the action of the stochastic driving force
associated to the heat bath, and 
${\cal X}[{\bf v}| f_{a,\mu},f_{b,\mu}]$ represents the flow of particles
of species $\mu$ between the two compartments.


The integral $I_{\mu\nu}$ describing collisions between
a particle of type $\mu$ and a particle of type $\nu$ can be represented as
\begin{equation}
I_{\mu\nu}[ {\bf v}_1| f_{a,\mu}, f_{a,\nu}]  = \sigma
\int d{\bf v}_2 \int' d\widehat{{\bf n}} 
(\widehat{{\bf n}}\cdot {\bf v}_{12})
\left( \frac{1}{r^2}
f_{a,\mu}({\bf v}_1'')f_{a,\nu}({\bf v}_2'') - 
f_{a,\mu}({\bf v}_1)f_{a,\nu}({\bf v}_2) \right)
\, ,
\end{equation}
where doubly primed symbols stand for pre-collisional
velocities and the $r^2$ factor in the denominator 
stems from inelasticity~\cite{vannoije}.

The number of particles of species $\mu$ in 
compartment $a$ is defined as
\begin{equation}
n_{a,\mu}(t) = \int d\br \int d\bv  f_{a,\mu}(\bv,t) \ .
\end{equation} 
Similarly, we define four partial granular temperatures:
\begin{equation}
n_{a,\mu} T_{a,\mu}(t) = 
\int d\br \int d\bv \frac{m_{\mu} v^2}{2} f_{a,\mu}(\bv,t) \ .
\end{equation} 
 
Now we derive the governing equations for occupation numbers
and kinetic temperatures. By integrating eq. (\ref{eq:bolt})
with respect to velocity ${\bf v_1}$, and recalling that both
the collision integral and the heat bath term
conserve the number of particles in each compartment, one can see that
the rate of change of $n_{a,\mu}$ is determined solely by the
exchange term ${\cal X}$:
\begin{equation}
\partial_t n_{a,\mu} (t) = 
\int d{\bf v}{\cal X}[\bv| f_{a,\mu},f_{b,\mu}] \,\, .
\label{eq:numero}
\end{equation}
According to the assumption of point 6 (see list above),
the exchange term has the form
\be
{\cal X}[\bv| f_{a,\mu},f_{b,\mu}]=-
\frac{1}{\tau_s}\theta(|\bv|-u_s)[f_{a,\mu}(\bv,t)-f_{b,\mu}(\bv,t)] \,\, .
\label{eq:boltzmann}
\ee
In other words, one assumes that particles having velocities larger
than a given threshold $u_s$ cross the barrier with probability
$\tau_s^{-1}$ per unit time.
The resulting equation for $n_{a,\mu}$ is
\be
\frac{{\rm d} n_{a,\mu}(t)}{{\rm d}t}=-\frac{V}{\tau_s}
\int{\rm d}\bv [f_{a,\mu}({\bf v},t)-
f_{b,\mu}({\bf v},t)]\theta(|{\bf v}|-u_s) \, .
\label{eq:salto}
\ee 
Mathematical convenience suggests an additional assumption for the
form of the heat bath operator ${\cal B}$:
we replace the effect of the periodic vibration of the piston
with a sequence of uncorrelated random kicks whose amplitude is
distributed according to a Gaussian~\cite{Puglisi2,Vibrated}.
The effect of this stochastic acceleration is described by the Langevin
equation
\begin{equation}
\frac{{\rm d}{\bf v}_i}{{\rm d} t}= \hat{\bf{\xi}_i} \, ,
\label{eq:due}
\end{equation}
where ${\bf {\xi}_i}$  
is a a Gaussian white noise with zero average, and variance given by
\begin{equation}
\langle {\xi}_{i}(t) {\xi}_{j}(t^\prime)\rangle =
\eta^2 \delta_{ij}\delta(t-t^\prime)
\label{drive} \, .
\end{equation}
Here $\eta^2$ measures the intensity of the stochastic driving.
This leads to the following forcing term:
\be
{\cal B}f_{a,\mu}=\frac{\eta^2}{2} 
\frac{\partial^2}{\partial v^2}f_{a,\mu}(\bv,t) \, .
\ee
Notice that particles receive an energy proportional
to their masses.
Next, we consider
the time evolution of partial kinetic energies $T_{\mu}$,
which is obtained by multiplying eq.(\ref{eq:bolt})
by $m_{\mu}\bv^2/{2}$ and integrating over velocity.
\begin{eqnarray}
\frac{1}{V}\partial_t (n_{a,\mu} T_{a,\mu}) = 
\frac{m_{\mu}}{2}
\sum_{\nu}  \int  d{\bf v} 
v^2 I_{\mu\nu}[ {\bf v} | 
f_{a,\mu}, 
f_{a,\nu}] + 
\frac{m_{\mu}}{2}\int  d{\bf v} 
v^2 {\cal B} f_{a,\mu} \nonumber\\
-\frac{m_{\mu}}{2\tau_s}
\int{\rm d}\bv v^2 [f_{a,\mu}({\bf v},t)-
f_{b,\mu}({\bf v},t)]\theta(|{\bf v}|-u_s),
 \qquad  
\end{eqnarray}


Finally, in order to derive an explicit expression for the governing
equations we assume Gaussian velocity distributions:
\be
f_{a,\mu}(\bv,t)=\frac{n_{a,\mu}}{V}\frac{m_{\mu}}
{2 \pi T_{a,\mu}}\exp \left (-\frac{m_{\mu}\bv^2}{2 T_{a,\mu}}\right) \, .
\label{eq:gauss}
\ee
Inserting eq. (\ref{eq:gauss}) into equation (\ref{eq:salto}) we obtain
\be
\frac{{\rm d} n_{a,\mu}(t)}{{\rm d}t}=\frac{1}{\tau_s} 
\left [n_{b,\mu} e^{-T_{\mu s}/T_{b,\mu}}-n_{a,\mu} 
e^{-T_{\mu s} /T_{a,\mu}} \right] \, ,
\label{eq:occupation}
\ee 
where the temperature $T_{\mu s}$ is given by $T_{\mu s}=\frac{1}{2} m_{\mu} 
u_s^2$ and the right-hand side
represents the difference between the incoming flux and the outgoing
flux of compartment $a$, as in the flux model.

The calculations involving the collision term 
are quite lengthy, but straightforward. Here we will not report the detailed
derivation, which can be found in refs.~\cite{Garzo,Equipart}.
The final form of the governing equation for the granular temperature
(say, of species 1 in compartment a) is:
\begin{eqnarray}
n_{a,1}(t)\frac{{\rm d} T_{a,1}(t)}{{\rm d}t}&=&
-\frac{2}{\tau_s}
\left[  n_{a,1} T_{a,1} e^{-T_{1 s}/T_{a,1}}-n_{b,1} T_{b,1} 
e^{-T_{1 s}/T_{b,1}}\right ]
\nonumber \\
&-&\frac{1}{\tau_s} \left[(n_{a,1} e^{-T_{1 s}/T_{a,1}}
-n_{b,1} e^{-T_{1 s}/T_{b,1}})
(2 T_{1 s}-T_{a,1})\right ]
\nonumber \\
&-&
\gamma_{11} n_{a,1} T_{a,1}-\gamma_{12}m_1 n_{a,1} 
\left(  \frac{T_{a,1}}{m_{1}}+\frac{T_{a,2}}{m_{2}}\right)
\nonumber \\
&-&\kappa_{12} m_1 n_{a,1} (\frac{T_{a,1}-T_{a,2}}{m_1+m_2})
+m_1\eta^2 n_{a,1} \,\, ,
\label{eq:energy}
\end{eqnarray}
The corresponding equations for other choices of species and compartment
are obtained by simple substitution of indices.
%
where the coefficients $\gamma_{\mu\nu}$ and $\kappa_{\mu\nu}$
are given by
\be
\gamma_{\mu\nu}=(1-r^2)\sigma\mu_{\nu\mu}^2
\frac{n_{a,\nu}}{ A}
\left( \frac{2T_{a,\mu}}{m_{\mu}} +  
\frac{2T_{a,\nu}}{m_{\nu}} \right)^{1/2} \, , 
\ee
\be
\kappa_{\mu\nu}=2 (1+r)
\sigma\mu_{\nu\mu}^2
\frac{n_{a,\nu}}{ A}
\left( \frac{2T_{a,\mu}}{m_{\mu}} +  
\frac{2T_{a,\nu}}{m_{\nu}} \right)^{1/2} \, .
\ee
The constant $A$ is the area available to the particles in each compartment. 
$\gamma_{\mu\nu}$ takes into account the effect of collisions between
particles belonging to species $\mu$ and $\nu$, respectively.
Interestingly, $\kappa_{\mu\nu}$ does not vanish in the
elastic limit ($r \to 1$), since it is associated with the 
restoring mechanism which tends to equalize the partial kinetic temperatures
of two different components in a standard fluid. 

The complete set of equations (\ref{eq:occupation}) and (\ref{eq:energy})
consists of eight non-linear coupled equations, of which only six are
independent, because the occupation numbers -- being globally conserved --
have to satisfy $n_{1,\mu}+n_{2,\mu}=\mathcal{N}_{\mu}$.

\section{The mean-field scenario}
To be able to compare the predictions of the proposed mean-field model
against the results of numerical simulation, we need to introduce
a suitable definition of $R$ (eq. \ref{R}) for the case of our thermal bath.
Specifically, we rewrite the first factor of equation (\ref{R}) as $mgh / (P^2/m)$,
where $P$ is the typical momentum transferred to a particle that collides with the piston.
We replace the energy scale $mgh$ -- associated to barrier crossing -- with the analogous
scale $T_{\mu s}$, and the transferred momentum scale $P$ with its equivalent for the
stochastic forcing of eq.~\ref{drive}, $m \eta \sqrt{\tau_s}$. Thus, for the mean-field model,
the analogous of eq.~\ref{R} can be written as:
\be
R_{MF}=\frac{T_{\mu s}}{m_{\mu}\eta^2\tau_s}(1-r^2) \frac{N^2 \sigma^2}{L_x^2} \, .
\label{RMF}
\ee
Obviously we expect the correspondence between $R$ and $R_{MF}$ to be useful only
for qualitative comparison of the mean-field model against numerical simulation.

We shall now consider some of the properties of the asymptotic solutions
of eqs.(\ref{eq:occupation}) and (\ref{eq:energy}).
The specific numerical values of the control parameters correspond approximately
to those used in the experiments by \cite{Mikkelsen}.

A symmetric solution,
characterized by equal state variables in the two compartments,
$n_{1,\mu}=n_{2,\mu}$ and $T_{1,\mu}=T_{2,\mu}$, always represents
a fixed point for the governing equations. 
However,
as the parameter $R_{MF}$ increases, 
such a symmetric fixed point becomes unstable. At small driving intensity or
high dissipation the solution ceases to be symmetric and the particles
of both species tend to
cluster in one of the two compartments. 
Such a symmetry breaking behavior
was studied experimentally and explained in terms of 
flux models \cite{Lambiotte,Mikkelsen}. By numerically integrating
eqs.(\ref{eq:occupation}) and (\ref{eq:energy}), we are also able to observe
the same simmetry-breaking behavior for the case of our mean-field description.

\begin{figure}[ht]
\vspace{-0.4truecm}
\includegraphics[angle=0,width=15.cm]{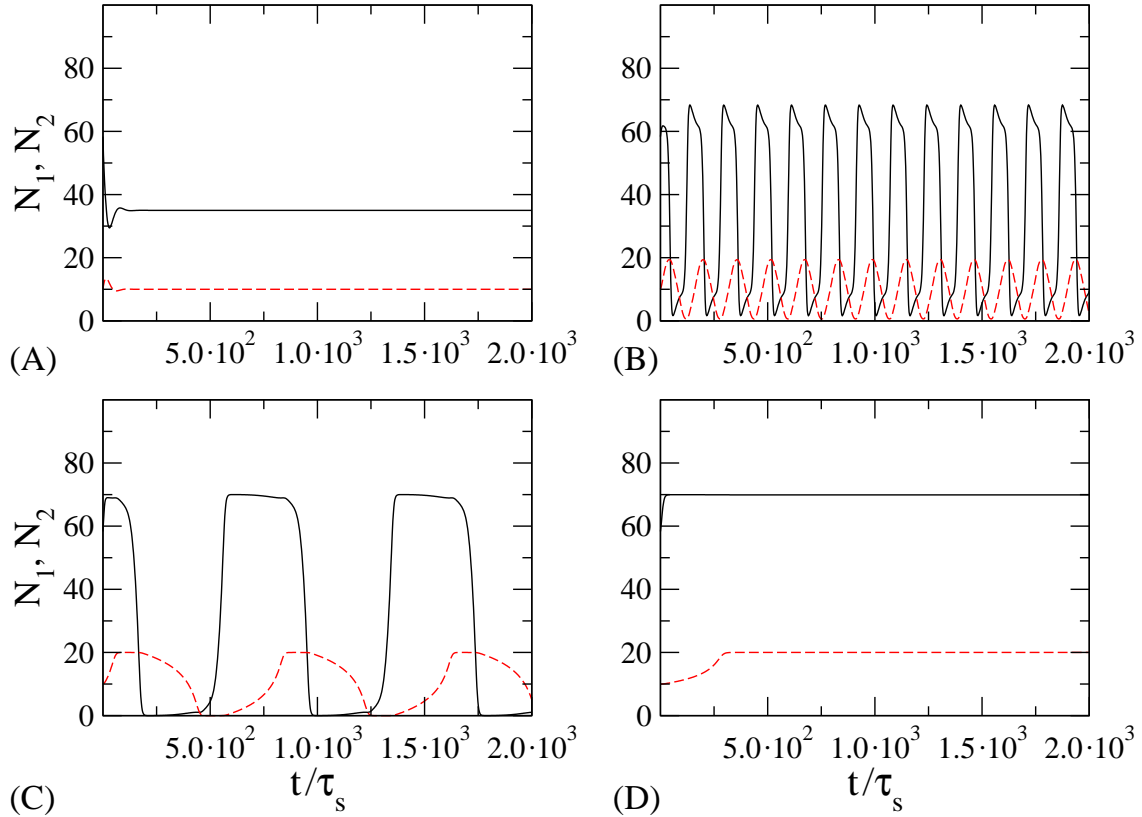}
\caption{{Four different behaviors obtained by varying the
dimensionless parameter $R_{MF}$, and keeping the mass ratio $m_2/m_1=8$ and the
population ratio $\mathcal{N}_1/\mathcal{N}_2=70/20$. The continuous line ($N_1$) refers to the light
species and the dotted line ($N_2$) to the heavy species, both in the right compartment.
In panel A ($R_{MF}=6.0$) the asymptotic solution is symmetric.
In panel B ($R_{MF}=12$) the occupation numbers oscillate in time.
Panel C ($R_{MF}=24$) shows a case where the solution displays 
oscillations with longer periods. Finally, panel D ($R_{MF}=48$)
illustrates a typical simmetry breaking solution, where occupation numbers
in the two compartments are different, for each species.
The time is measured in units $\tau_s$.}}
\label{fig:mft_solut}
\end{figure}

In fig.~\ref{fig:mft_solut}
we show the temporal evolution of occupation numbers for the two species
(in a single compartment) for different choices of the 
driving intensity. We observe two dynamical scenarios. In the first scenario,
the occupation numbers approach a stationary symmetry-broken solution
via a simple bifurcation (i.e a critical point).
This corresponds to the situation
considered in ref.~\cite{Ioepuglio}, and occurs for small mass differences
or small concentrations ($m_2/m_1=2$, for example). In the second scenario,
the temporal evolution of the occupation numbers approaches a limit cycle
so that that particles of each species oscillate back and forth between compartments.
This occurs when the difference in mass is large, or when masses are equal
but sizes are different enough, in analogy to the observations
of Lambiotte et al.~\cite{Lambiotte}.

\begin{figure}[ht]
\includegraphics[angle=0,width=12.cm]{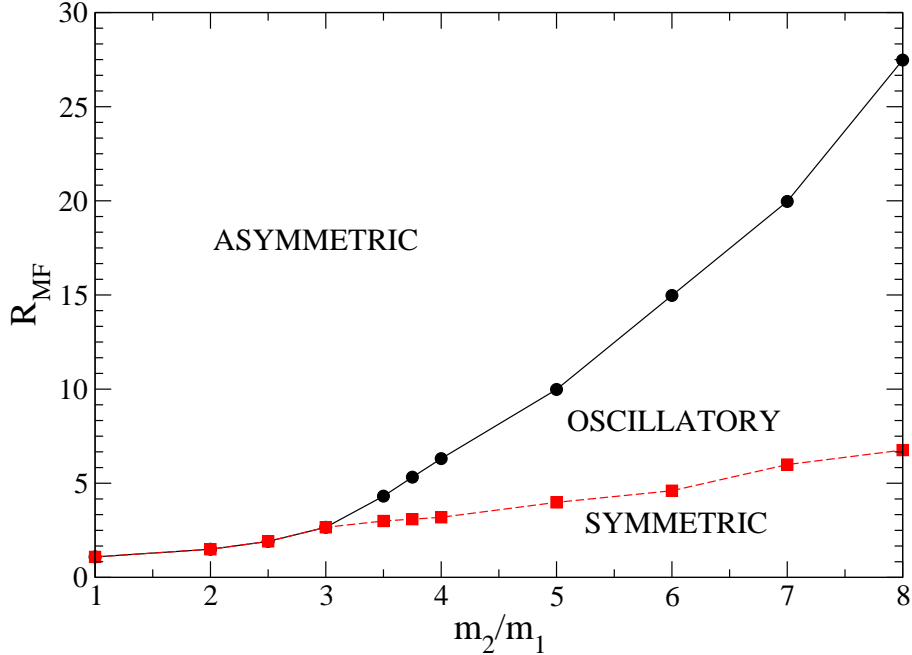}
\caption{\small{Phase diagram for a system with $\mathcal{N}_1=70$ light
particles and $\mathcal{N}_2=20$ heavy particles. The transition values
of $R_{MF}$  are plotted as functions of the mass ratio $m_2/m_1$.}}
\label{phase1}
\end{figure}

Although the system is far from equilibrium and has a finite extension
we shall employ the term ``phase diagram'' in order to stress the existence
of different dynamic regimes.
The resulting phase diagram in the plane 
$R_{MF}$ vs. $m_2/m_1$
is shown in fig.~\ref{phase1}. No oscillations can be observed for
small mass asymmetry ($\frac{m1}{m2} \simeq 1$) 
and the crossover from the symmetric phase to the asymmetric phase
is similar to what occurs in the case of a one-component system.
However, when the mass asymmetry increases ($m_2/m_1 \gtrsim 3$) 
an intermediate oscillatory regime appears,
and no stationary solution is attained anymore.

\begin{figure}[ht]
\includegraphics[angle=0,width=12.cm]{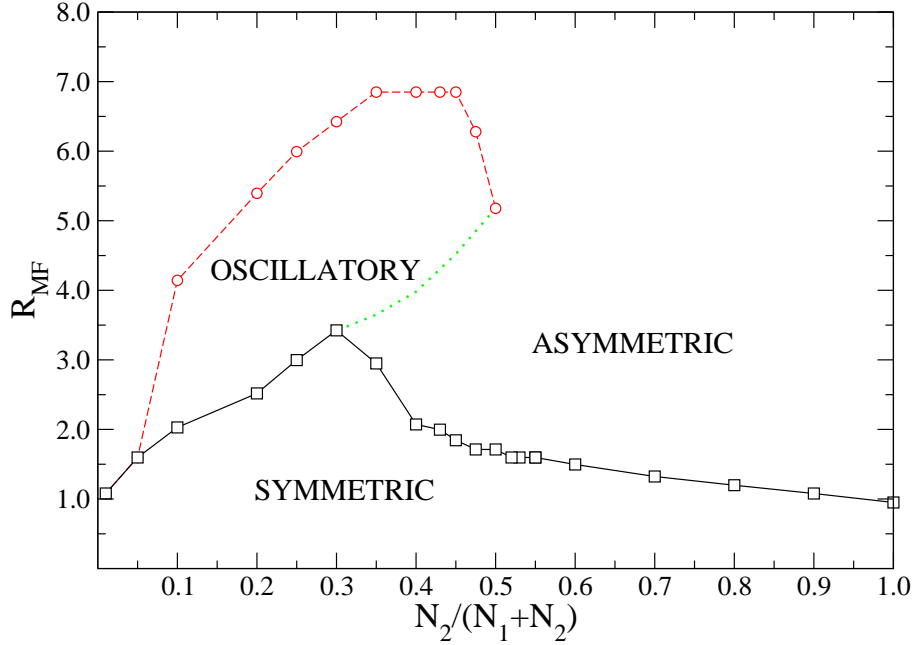}
\caption{\small{Phase diagram for a system with $\mathcal{N}_1+\mathcal{N}_2=90$
particles as function of $\mathcal{N}_2/(\mathcal{N}_1+\mathcal{N}_2)$, for a mass
ratio $m_2/m_1=4$.}}
\label{phase2}
\end{figure}

The effect of concentration on the appearance of oscillations
is instead shown in fig.~\ref{phase2}. For small concentrations
of the heavy species $\mathcal{N}_2/(\mathcal{N}_1+\mathcal{N}_2)$ 
there is a direct crossover, as  $R$ decreases, 
from the asymmetric ``phase'' to the symmetric ``phase''. 
When the concentration $\mathcal{N}_2/(\mathcal{N}_1+\mathcal{N}_2)$ of the heavy particles
increases, there appears an island of the oscillating ``phase'', which 
subsequently disappears when the concentration of heavy particles becomes too
large. The oscillation period decreases as $R_{MF}$ decreases
and is larger near the boundary between the ``broken phase''
and the ``oscillatory phase''. In the following, in order to simplify the notation, we shall denote 
$N_1=n_{1,1}$ and $N_2=n_{2,1}$.



\section{Numerical experiments}

In order to verify whether the hard disk model correctly describes
experimental observations \cite{Lohse} and also to 
probe the qualitative validity of our mean-field scenario,
we now turn to discuss a set of numerical experiments.
We confirm that the behavior of the mixture, as already revealed 
by mean-field calculations, strongly depends on the specific choice
of control parameters.

\begin{figure}[ht]
\begin{center}
\includegraphics[angle=0,width=\columnwidth]{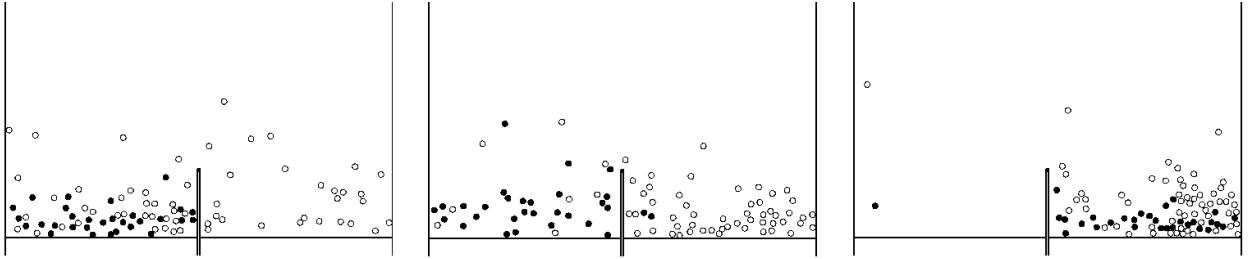}
\end{center}
\caption{\small{
Typical snapshots of the process described in Section V. Open circles 
indicate the light particles, black circles
the heavy particles. System parameters are $\nu=47$ Hz, $L_x=34 \sigma$, $m_2/m_1=8$, 
$\mathcal{N}_1=60$, $\mathcal{N}_2=30$, $A=0.4 \sigma$ and $\sigma=0.5 cm$. }}
\label{snapshot}
\end{figure}

The simulated dynamics consists of a succession of ballistic
trajectories and inter-particle or wall-particle collisions. 
With respect to the code employed previously
(see references~\cite{Daniela} and~\cite{convection}) we introduce
a dividing, elastic barrier, visibile in fig. \ref{snapshot}.
The two species are chosen to be smooth rigid disks of equal diameters
$\sigma=0.5$ cm and unequal masses $m_1=6.545\cdot 10^{-2}g$
and $m_2=5.236\cdot 10^{-1}g$, respectively. 
The mixture is composed of $\mathcal{N}_1=60$ light particles and $\mathcal{N}_2=30$ heavy 
particles. 
The vibrating base is driven according to a symmetric sawtooth waveform
and the vibration amplitude is set to $A=0.4 \sigma$.
Finally, the width $L_x$ of the compartment ranges from $34 \sigma$ to
$240 \sigma$, while its vertical dimension is $L_z=82.4 \sigma$,
and  the height of the dividing wall is $h = 12 \sigma$.

Our observables are computed as instantaneous averages 
in the case of global quantities, such as the occupation numbers for
the two species and the partial kinetic energy per particle, and
as time averages (ergodic averages) for other quantities, such as 
densities, temperatures and velocity distribution functions.

\begin{figure}[ht]
\begin{center}
\includegraphics[angle=0,width=\textwidth]{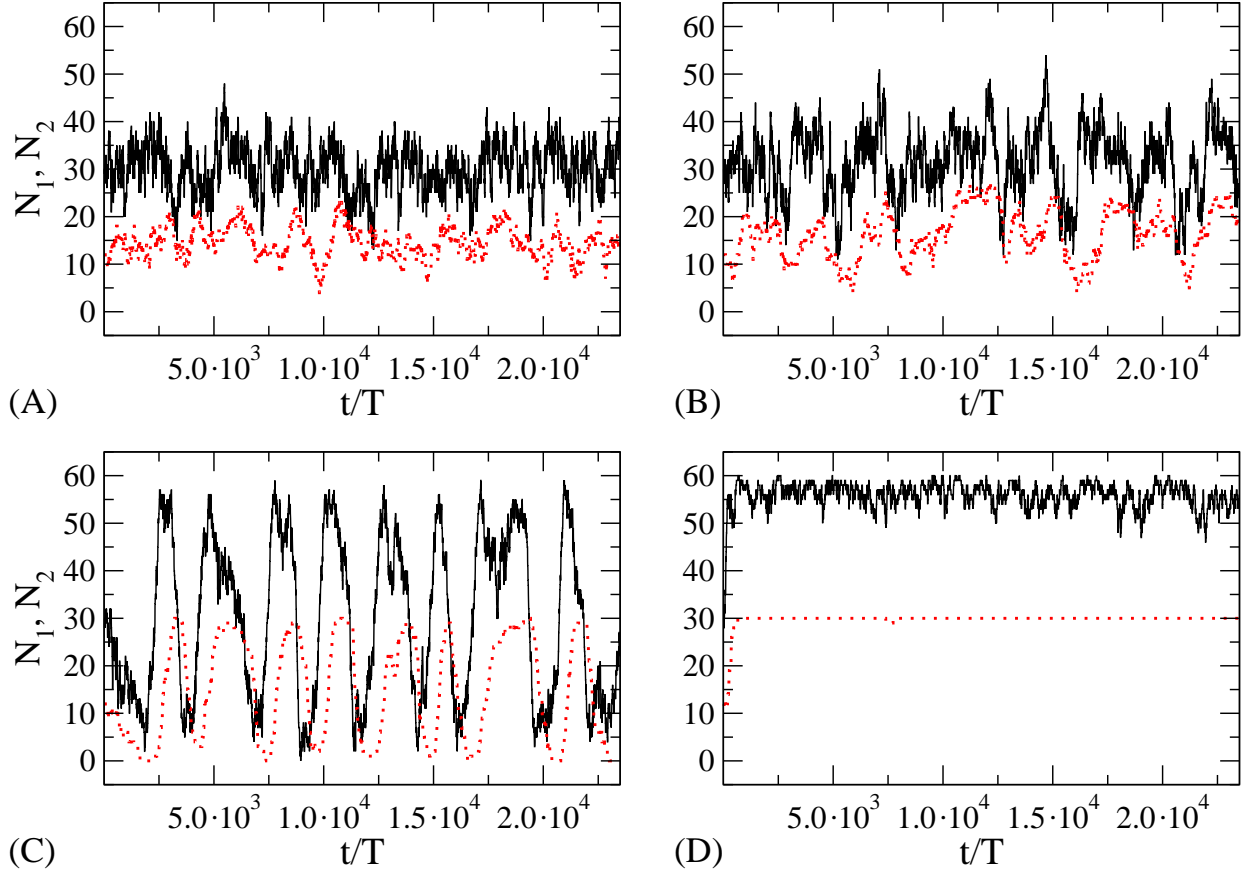}
\end{center}
\vspace{-0.7truecm}
\caption{
Number of light particles (solid line, $N_1$) and heavy particles
(dotted line, $N_2$) in the left compartment, as a function of time,
for different values of $L_x$: $L_x=160 \sigma$ (A), $L_x=120 \sigma$ (B),
$L_x=80 \sigma$ (C) and $L_x=62 \sigma$ (D).
System parameters: $r=0.85$, $\nu=47$ Hz
(period $T = 1/\nu$), $A=0.4\sigma$, $\sigma=0.5 cm$
and $m_2/m_1=8$.
The panels correspond to $R=5.8$ (A), $10.4$ (B), $23.4$ (C), $39.0$ (D).}
\label{panels4}
\end{figure}

For strong driving (i.e. small $R$, as already shown in the mean-field case)
the system approaches a stationary state where the number of particles
in each compartment, or average, is the same.
Such a steady state is maintained by a continuous exchange of
particles of both species between the compartments. The average
kinetic energies
are sufficiently high so that particles can jump over the barrier.
Upon increasing $R$ the symmetry is spontaneously broken: the system 
can either attain an asymmetric stationary state, or approach a limit cycle
where populations keep switching back and forth between compartments
in a coherent fashion. These scenarios are shown
in figures \ref{panels4} for $m_2/m_1=8$, $r=0.85$ and $\nu=47$ Hz,
where the crossover from one regime to the other occurs by varying
the value of $L_x$.
For $L_x=160 \sigma$ and $L_x=120 \sigma$ the particles are nearly
equidistributed among the compartments. 
On the other hand, when the system is narrow enough ($L_x=62 \sigma$)
the particles -- after an initial transient -- tend to localize
in the same compartment for time intervals longer than our observation time.
However, when we change the width of the compartment from $L_x=62 \sigma$
to $L_x=80 \sigma$, pronounced oscillations appear and the populations appear
to move back and forth. The global picture is consistent with our previous
mean-field analysis.

\begin{figure}[ht]
\begin{center}
\includegraphics[angle=0,width=\textwidth]{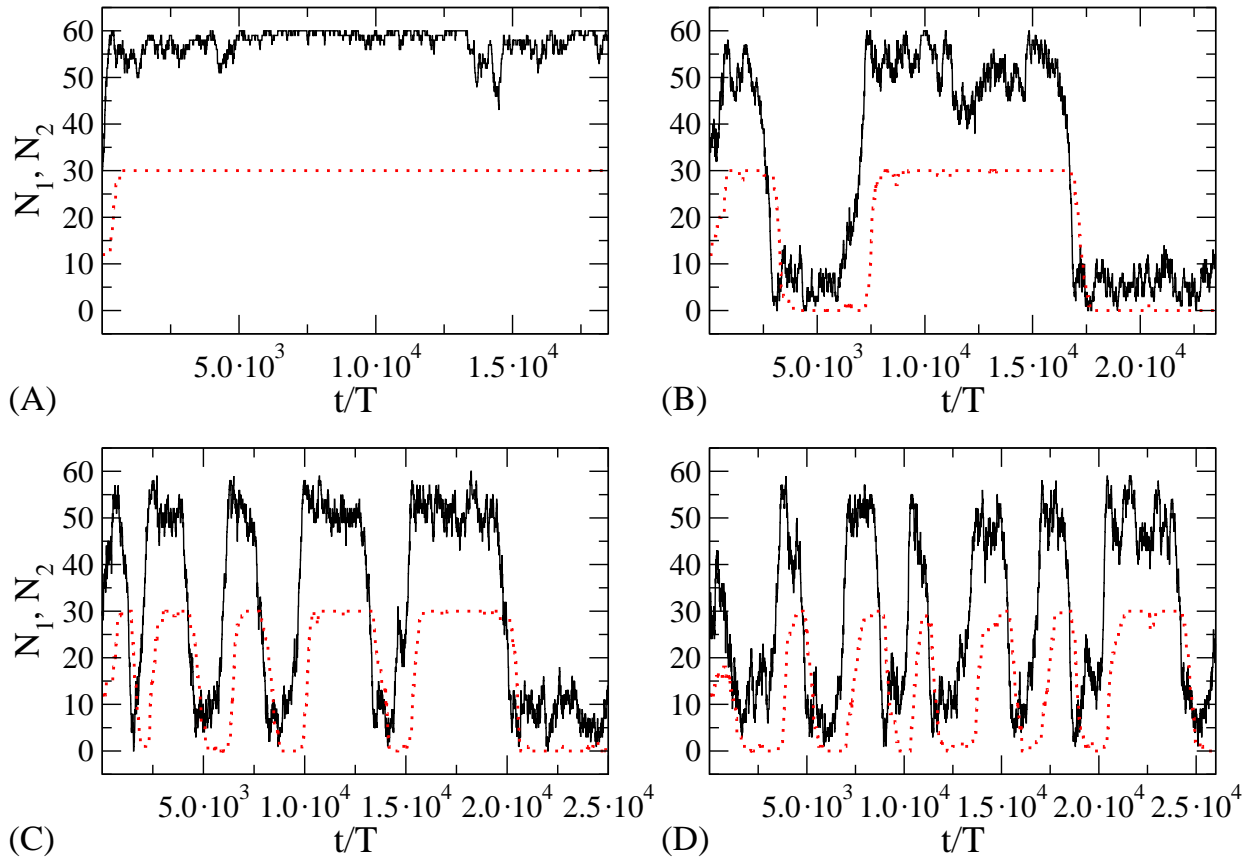}
\end{center}
\vspace{-0.7truecm}
\caption{
Number of light particles (solid line, $N_1$) and heavy particles
(dotted line, $N_2$) in the left compartment, as a function of time,
for different driving frequencies $\nu$ (period $T=1/\nu$):
$\nu =37$ Hz (A), $\nu =47$ Hz (B),
$\nu =50$ Hz (C), $\nu =52$ Hz (D).
System parameters: $r=0.85$, $ L_x=68 \sigma$, $A=0.4\sigma$,
$\sigma=0.5~cm$ and $m_1/m_2=8$.
}
\label{panels4-freq}
\end{figure}

\begin{figure}[ht]
\begin{center}
\includegraphics[angle=0,width=\textwidth]{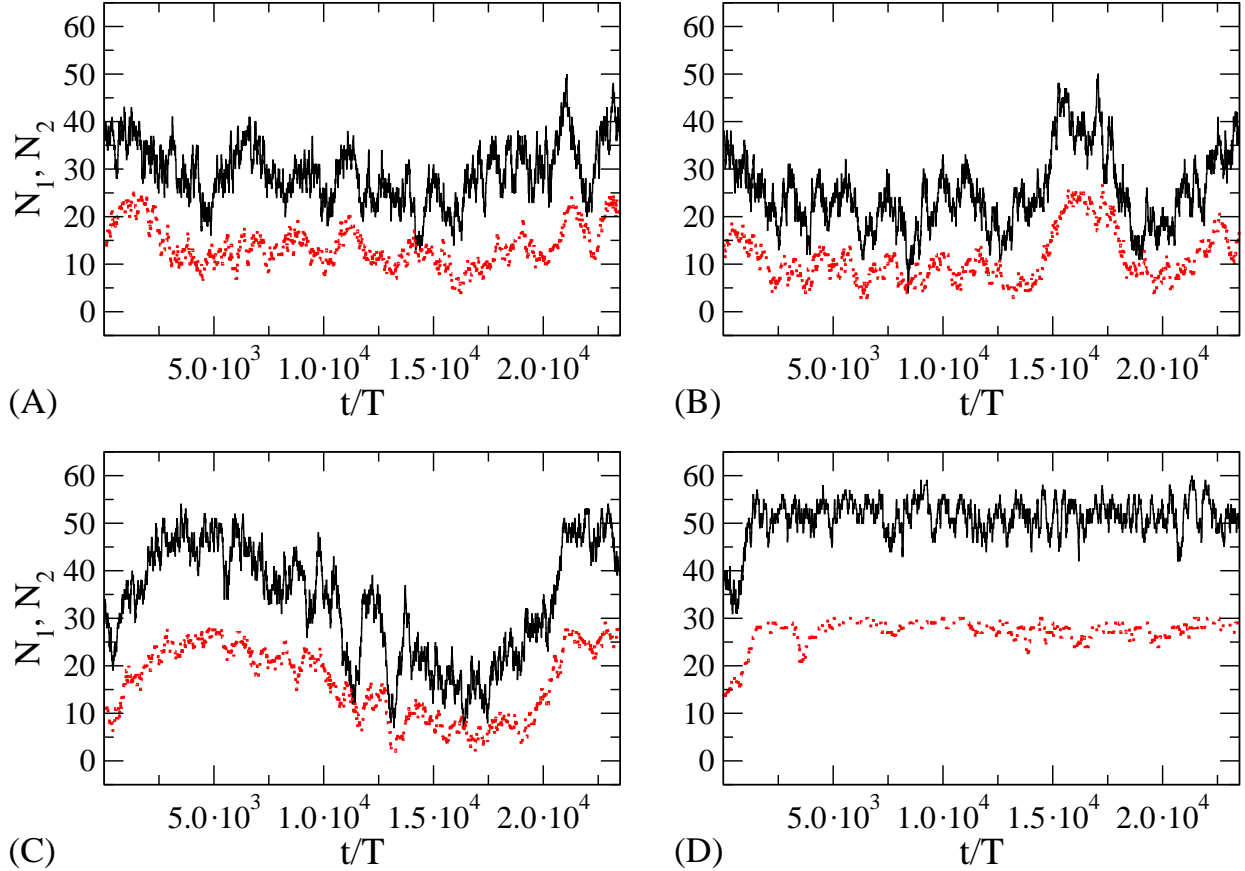}
\end{center}
\vspace{-0.7truecm}
\caption{\small{
Number of light particles (solid line, $N_1$) and heavy particles
(dotted line, $N_2$) particle in the left compartment, as a function
of time, for different choices of $L_x$: $L_x=240 \sigma$ (A),
$L_x=220 \sigma$ (B), $L_x=200 \sigma$ (C) and $L_x=160 \sigma$ (D).
System parameters: $\nu=47$ Hz, $A=0.4\sigma$, $\sigma=0.5 cm$ 
and $m_2/m_1=2$.
The panels correspond to $R=2.6$ (A), $3.1$ (B), $3.7$ (C), $5.8$ (D).}}
\label{panels4dp}
\end{figure}

We also explore the effect of varying the driving frequency
at fixed $L_x$, and we find that the breakdown of the symmetric
stationary state still occurs, as shown in fig.~\ref{panels4-freq}.
Moreover, the oscillatory behavior only appears for large mass asymmetry.
On the other hand, the time series of fig.~\ref{panels4dp} shows that the
corresponding mass asymmetry ($m_2/m_1=2$) is not large enough to induce
clean oscillations, so that only the phenomenology of 
one-component driven granular gases can be observed. 

Finally, we remark that the same kind of oscillations  may occur
when the diameters of the two species are different enough,
even in the case of equal masses (data not shown). 


Let us consider in more detail the oscillatory behavior we observe
(for example, refer to panel C of figure \ref{panels4-freq}).
After an initial clustering of both species in a
compartment -- say the left -- a net rightward flux of light particles
establishes and persists until a sufficient number of them have changed
compartment. At this point the heavy particles, too, start jumping to the
right, eventually creating in the right compartment a cluster of both
species which is totally similar to the initial situation of the left
compartment. After reaching this stage, the process repeats itself
in the opposite direction.

Such a peculiar behavior has its origin in the mass and/or size
difference between the two species
and in the associated breakdown of energy equipartition
that occurs in a binary vibrated granular gas~\cite{Feitosa,Pagnani,Ciroart}. 
Krouskop and Talbot~\cite{Talbot} studied how the mass ratio
affects the exchange of energy of each species and computed 
the average energy change
of a particle of species $\mu$ colliding with a particle of species $\nu$.
Assuming that each component has a Maxwell 
velocity distribution and a partial temperature 
$T_{\mu}$, the average energy change in 2D is
\be
\Delta E_{\mu\nu} = \mu_{\mu\nu}\frac{(1+r)}{2\sqrt{2}}\biggl[
(1+r)\frac{T_\nu m_\mu+T_\mu m_\nu}{m_\mu(m_\mu+m_\nu)}-
2\frac{T_\mu}{m_\mu} \biggr] \, .
\label{f21}
\ee  
Such an equation shows that for $T_2/T_1>1$ a light particle -- on average --
gains energy on colliding with a heavy particle.
In our case, the imbalance in temperatures is guaranteed by the fact
that heavy particles receive more energy than light ones when colliding
with the vibrating base. Such an imbalance is roughly proportional
to the mass ratio. Thus heavy particles are able to gain more energy
from the driving base, and they can also tranfer it to light particles,
effectively heating them.
Since the system is not in equilibrium, kinetic
temperatures depend on the particle numbers in each compartment 
and are not known {\it a priori}.

\begin{figure}[ht]
\begin{center}
\includegraphics[angle=0,width=\columnwidth]{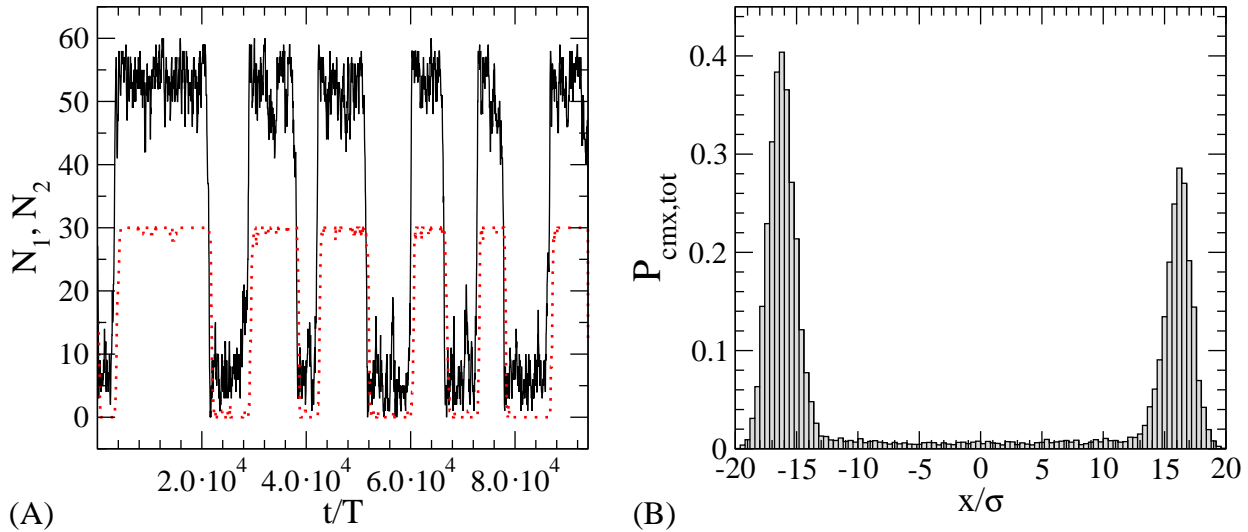}
\end{center}
\vspace{-0.7truecm}
\caption{\small{ Panel A: time evolution of the 
occupation numbers $N_1$ and $N_2$ in the left compartment.
Panel B: probability distribution of the horizontal position $x$
of the center of mass (whole system). The left-right asymmetry
is due to the limited size of the sample used to compute the histogram.
System parameters are $\nu=47$ Hz, $A=0.8\sigma$, 
$L_x=68 \sigma$ and $m_2/m_1=8$.}}
\label{distribu}
\end{figure}

\begin{figure}[ht]
\begin{center}
\includegraphics[angle=0,width=11.cm]{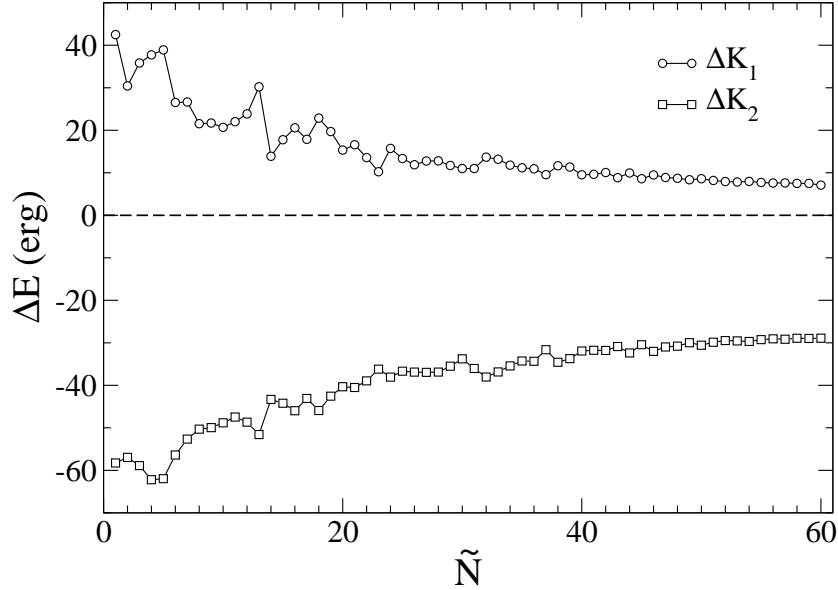}
\end{center}
\vspace{-0.7truecm}
\caption{Average energy change for particle-particle collisions,
as a function of the number of light particles $\tilde{N}$ present
in the compartment where the majority of the heavy 
particles is located.
$\Delta K_1$ (open circles) and 
$\Delta K_2$ (squares) correspond
to the energy change for light and heavy particles respectively,
in the full compartment.
\label{figdeltaE}}
\end{figure}

In order to check the validity of the above argument, we focused on the
steady-state dynamics between transitions, and numerically evaluated
the partial temperatures $T_1$ and $T_2$ in the compartment hosting
the majority of heavy particles. Such temperatures mainly depend
on the number $\tilde{N}$ of light particles, while they are not
signifcantly affected by the number of heavy particles, because the
latter population is basically constant between transitions
(see fig.~\ref{distribu}).
On inserting the above partial temperatures into equation (\ref{f21}),
and weighting the energy changes with the numerically evaluated
collision frequencies, we obtain the average energy changes per particle
and per collision, separately for each species:
\begin{equation}
\Delta K_1=\frac{\Delta E_{11} M_{11}+\Delta E_{12} M_{12}}{M_{11}+M_{12}}
\end{equation}
 and
\begin{equation}
\Delta K_2=\frac{\Delta E_{21} M_{12}+\Delta E_{22}M_{22}}{M_{22}+M_{12}} \,\, .
\end{equation} 
Here $M_{\mu \nu}$ represents
the number of collisions per unit time between species $\mu$ and $\nu$.
Fig.~\ref{figdeltaE} shows $\Delta K_1$ and $\Delta K_2$ as a function
of ${\tilde N}$. We observe that light particles, on average, gain energy.
Such a gain increases as ${\tilde N}$ decreases, because in this situation
$M_{12}$ is much larger than $M_{11}$. Thus light particles are more likely
to jump into the empty compartment, leaving the heavy ones behind.
At this point the heavy particles -- no longer transferring energy
to the other species -- become more energetic and also move to the other
compartment, joining the light particles once more.

Let us now consider in more detail the statistical properties of 
the system by analyzing a particular case
shown in fig.~\ref{distribu}A. One can observe 
that the occupation numbers fluctuate around
well defined plateau values for periods as long as several thousands
driving cycles (of duration $T \equiv 1/\nu$).

Interestingly, the distribution of the horizontal position of the 
center of mass displays a double peak structure, indicating
that the system spends most of the time in configurations where the majority
of the particles is on one side -- an asymmetric situation.

\begin{figure}[ht]
\begin{center}
\includegraphics[angle=0,width=12cm]{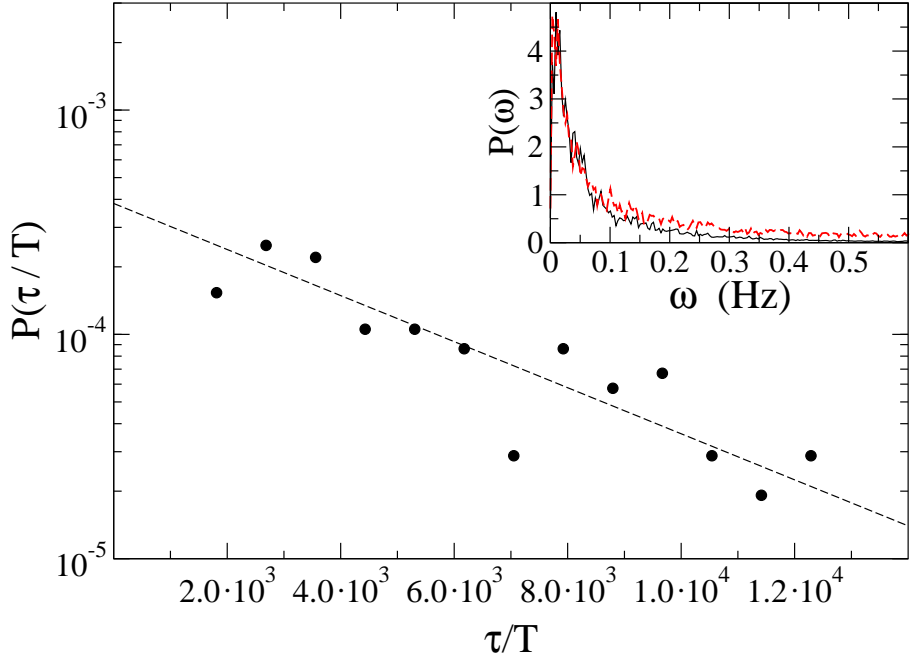}
\end{center}
\vspace{-0.7truecm}
\caption{\small{Probability distribution of the residence time $\tau$
for the (hardened) occupation number $N_2(t)$ of heavy particles. 
The dashed line is an exponential fit.
In inset: power spectrum $P(\omega)$ of the hardened occupation number
(solid line) and power spectrum of the synthetic dichotomic signal
used for comparison in the text (dashed line).
Simulation parameters are the same as in fig.~\ref{distribu}.
}}
\label{figresidence}
\end{figure}

We can better study the switching dynamics of fig.\ref{distribu}A
by defining a collective residence time $\tau$ as the period 
during which at least $80$\% of the heavy particles persist
in a given compartment. The ensuing distribution of residence times
is shown in fig.~\ref{figresidence}.
The average residence time is about $6 \cdot 10^3 T$ and the distribution
appears to be exponential.
In order to ascertain whether the
oscillations are periodic, as reported by Lambiotte et al. \cite{Lambiotte}
and predicted by our mean-field model, we study the signal
of fig.~\ref{distribu} in the frequency domain. The resulting power spectrum
$P(\omega)$ is shown in the inset of fig.~\ref{figresidence}
and displays no evidence of periodicity.
As a further check, we generated a ``synthetic'' bistable signal
by assuming that the occupation numbers of the heavy species
in each compartment can assume only two values (say $N_2$ and $0$),
and that the residence times are randomly distributed according
to an exponential density distribution whose characteristic time is
the same as the one we measured. The corresponding power spectrum
is also shown in inset of fig.~\ref{figresidence},
and appears to be very close to the original one.
Thus the numerically observed bistability is not periodic,
but has a stochastic nature. Such a stochastic switching
resembles what occurs in thermally activated
processes~\cite{dererumpallettarum}.
Our mean-field model is of course unable to reproduce non-deterministic
features, but it can still account for the non-stationary, bistable
character of the reported dynamics.


\begin{figure}[ht]
\begin{center}
\includegraphics[angle=0,width=13cm]{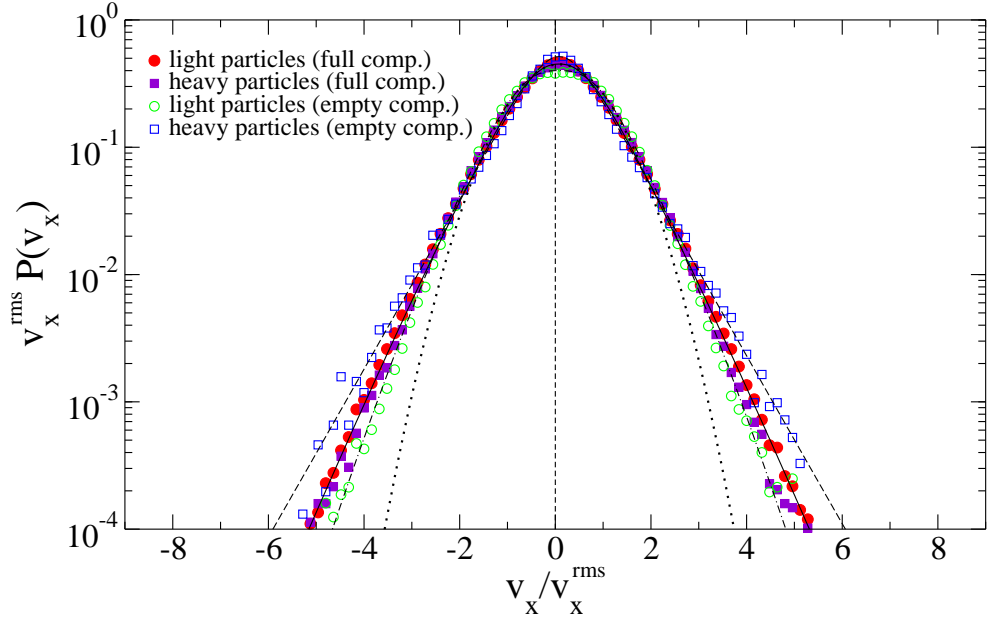}
\end{center}
\vspace{-0.7truecm}
\caption{\small{The rescaled 
velocity distribution functions $v_x^{rms}P(v_x)=f(v_x/v_x^{rms})$ 
for horizontal velocities.
The open and closed symbols correspond to the ``empty'' and ``full'' compartment, respectively.
Circles and the
squares correspond to the light and heavy species, respectively. 
The independent variable $v_x$ was rescaled by its mean
squared value $v_x^{rms}$: $38.29 cm/s$ (component 1) and $22.12 cm/s$ (component 2) for the ``full''
compartment; $59.44 cm/s$ (component 1) and $32.40 cm/s$ (component 2) for the ``empty'' compartment.
The solid and dash-dotted lines are the fitting laws discussed in the text,
while the dotted curve is a Gaussian distribution, as an aid for the eye.}}
\label{figpdv}
\end{figure}

Since the fast particles play a major role in activated processes,
we computed the velocity probability distributions for the two species,
in each compartment. Such distributions are shown in fig.~\ref{figpdv},
for the horizontal component of the velocity. On rescaling by the corresponding
mean squared velocities, they appear to collapse on two curves,
one for the ``empty'' compartment and another for the ``full'' one.
The two distributions deviate from a Gaussian law and can be fitted by
$f(v_x)=A/(exp(\beta v_x^{\lambda})+exp(-\beta v_x^{\lambda}))$ (as already found
in ref~\cite{Daniela}) with the following values for the fitting parameters:
$A_f=0.9,~\beta_f=1.38,~\lambda_f=1.14$ for both components in the ``full'' compartment and
$A_e^{(1)}=0.8,~\beta_e^{(1)}=1.1,~\lambda_e^{(1)}=1.35$ and
$A_e^{(2)}=1.03,~\beta_e^{(2)}=1.6,~\lambda_e^{(2)}=0.98$ for the component
1 and 2, respectively, in the ``empty''
one.
It is interesting to observe that the exponent $\lambda$ increases
with the occupation number, in agreement with the idea that high velocity tails
are associated with those particle that underwent a smaller number
of inter-particle collisions.

\begin{figure}[ht]
\begin{center}
\includegraphics[angle=0,width=13cm]{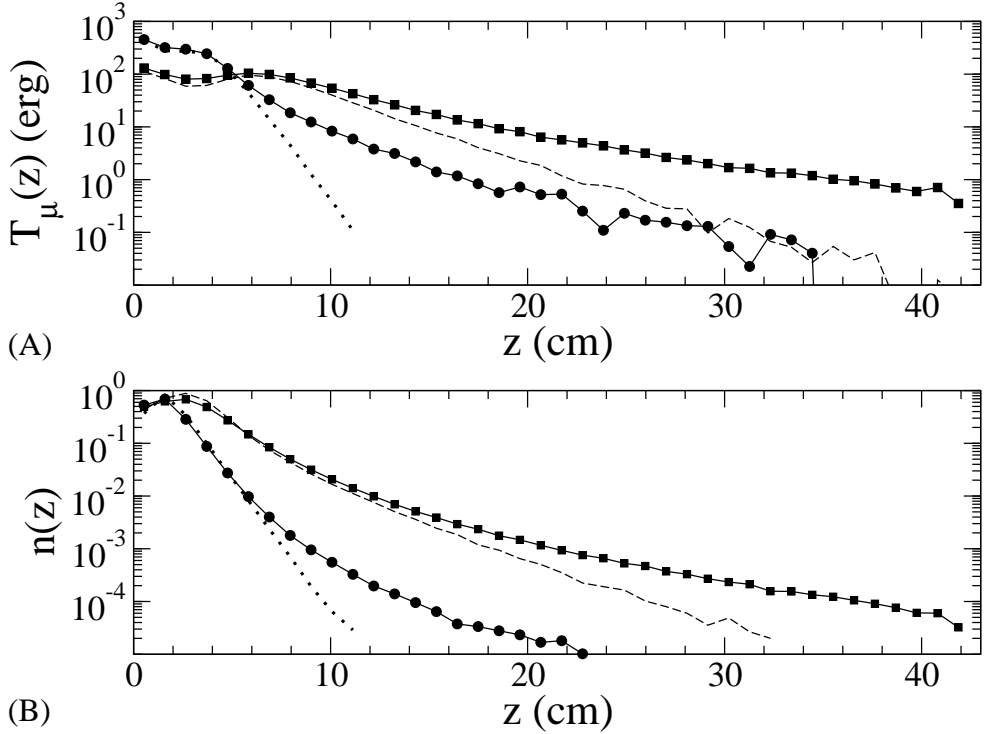}
\end{center}
\vspace{-0.7truecm}
\caption{\small{The granular temperature (a) and density (b) 
profiles versus height, for the ``full'' compartment. Square and 
circles correspond to light and heavy components, respectively,
for the case with two compartments. The dashed and the dotted lines correspond
to light and heavy components in the case of a single compartment.
}}
\label{figprofili}
\end{figure}

We now turn to consider the vertical temperature and density profiles
(see fig.~\ref{figprofili}).
The temperature profiles indicate that the kinetic temperature
of the heavy species is larger than that of the light one
(being the values of these two quantities dominated by the
bulk of the distributions). Remarkably, in the upper region the light
particles are more energetic than the heavy ones.

The density profiles show peaks at different positions for the two species, 
indicating that the light
particles tend to stay above the heavy ones. 
One can also observe that
the upper region, well above the top of the
dividing barrier, contains mostly light grains.

For the sake of comparison,
we also performed simulations of a reference system with
a single compartment \cite{Meerson}. For small values of $z$
the density profiles $n_1(z)$
and $n_2(z)$ result very similar to the case of the two-compartment system.
For higher values of $z$, both $n_1(z)$ and $n_2(z)$
are larger than the corresponding profiles in the reference system.
The local granular temperatures $T_1(z)$ and $T_2(z)$
display a similar behavior.
For small values of $z$, $T_2(z)$ is larger that $T_1(z)$,
because heavy grains gain a larger amount
of energy when colliding with the vibrating base,
but for large values of $z$ this relation is reversed:
light particles are ``hotter'' and their temperature profile is
characterized by a slower exponential decay.
This is consistent with the mechanisms discussed in section V.

\section{Conclusions}

In this paper we studied the behavior of a two-dimensional, driven,
inelastic granular mixture in a compartmentalized container. Our work is
hinged on two complementary approaches.
The first approach consists in reducing the dynamics,
by means of an appropriate coarse-graining procedure,
to a simple set of non-linear coupled ordinary differential equations
for the relevant observables, namely the number of particles 
and the average kinetic energies in each compartment. These mean-field
equations are solved numerically and show the existence of three qualitatively
different regimes: 1) for ``low'' values of the dimensionless parameter $R_{MF}$
the asymptotic state of the system is symmetric, i.e. the particles of each
species are equidistributed in the two compartments; 2) for intermediate
values of $R_{MF}$ the system attains a limit cycle where the populations
in each compartment exhibit an oscillatory behavior, and segregation
of species is observed; 3) for high values of $R_{MF}$ the system approaches
an asymptotically steady state, with unequal populations in the two
compartments and no segregation of species.

The second approach is a more realistic event-driven simulation with physical
parameters similar to those employed in laboratory experiments.
Our simulations show qualitatively similar scenarios.
However, whereas in the mean-field description 
we find two stationary regimes, a symmetric configuration and 
an asymmetric one,
separated by an oscillatory regime, our numerics
indicate that the intermediate oscillatory ``phase'' is indeed
a stochastic, bistable regime.
In both cases, the bistability emerges only when the masses or diameters
of the species are different enough. In addition to that, the typical time
scale characterizing transitions in the bistable regime grows as one departs
from the asymmetric phase and moves towards the symmetric phase. 
The mechanism that sustains such a non-stationary dynamics
was explained in terms of the microscopic kinetics of a mixture of inelastic 
particles.


\section{Acknowledgments}
We wish to thank Renaud Lambiotte for sending us a preprint
concerning the behavior of a granular mixture of grains of different
sizes. We also thank G. Mari for having participated to the early
stages of the present work.
U.M.B.M. acknowledges the support of the 
Project Complex Systems and Many-Body Problems
Cofin-MIUR 2003 prot. 2003020230.

\end{document}